# On Coaxial Microcalorimeter Calibration

L. Brunetti[1] , L. Oberto[1,2,*]

[1]Istituto Nazionale di Ricerca Metrologica – Strada delle Cacce 91, 10135, Torino, Italy
[2]Politecnico di Torino – Corso Duca degli Abruzzi 24, 10129, Torino, Italy

**Abstract**
Primary power standards in the microwave domain are realized using a calorimetric technique, usually identified with the used measurement system, i.e., the microcalorimeter. It is adjusted for measurement of power ratios with a relative accuracy that, after an appropriate system calibration, is of order of $10^{-3}$, at least in the microwave domain (1 GHz-18 GHz). Hereby we describe the calibration process implemented at the Istituto Nazionale di Ricerca Metrologica (Italy) for realizing a coaxial power standard based on indirect heating thermocouples. Particular regard is devoted to describe the nearly ideal thermal load used for determining the microcalorimeter losses and their influence on the measurand accuracy.

## 1 – Introduction

For intrinsic properties of the electromagnetic field, in general not conservative, electromagnetic power is a quantity very important from the metrological point of view because it is, operatively, always a well defined, even at high frequency (HF), conversely, e.g., the voltage [1]. Consequently, many National Metrology Institutes (NMIs) realize the HF primary power standards through an experiments, which the microcalorimetric technique plays the fundamental role in [2 - 6].

Originally, the microcalorimeter technique, in the following microcalorimeter, was developed for sensor based on bolometers, which the dc-substitution method [7] applies to. This method allows tracing the HF power standard to the direct current (dc), a SI quantity.

Basically, the microcalorimeter allows the HF loss measurement of a bolometer mount, that is, the measurement of its *effective efficiency* $h_e$, a parameter that implicitly defines the *power standard*.

Bolometers are still very popular in the primary laboratories in spite of some drawbacks. They are very sensitive to absolute temperature variations, have a limited dynamic range (about from 0.1 mW to 10 mW only), are downward frequency limited below 10 MHz because of an internal decoupling capacitance and, finally, they are available on the market with discontinuity because used in only few commercial applications. By the early 1990s, the Physikalisch Technische Bundesanstalt (PTB, Germany) proposed an alternative to the bolometric sensors based on indirect heating thermocouples [8]. This device is another true RMS power sensor that can still be measured and traced to the dc standard as the bolometric sensors but without their limitations.

The Istituto Nazionale di Ricerca Metrologica (INRiM) went further the PTB proposal and realized a new twin-line dry Microcalorimeter optimized for using this kind of sensors [9, 10]. This system allowed an extension of the metrological capabilities of the INRiM in HF field. Indeed, power standard is presently available with continuity from dc to 26.5 GHz and it can be extended to 40 GHz if we operate with 2.92 mm coaxial transmission line.

After these historical notes, we come to consider the main contribution given by INRiM to microcalorimeter improvements. The power standard accuracy depends on the ability to evaluate the losses of the microcalorimeter HF feeding lines, which constitute the main error source since ever. These losses could be calculated through complicated simulation processes but, more efficiently, they are measured if a sensor mount is available of known $h_e$. Anyway, there is another possibility based on the use of a full reflecting termination. In describing the microcalorimeter calibration we explain how a nearly ideal calorimeter load is realized having $h_e$ equal to 1 or of well measurable value.

---

[*] Corresponding author: Luca Oberto, l.oberto@inrim.it

**2 – Power standard rationale**

When HF power is supplied to a power sensor, e.g. a thermoelectric sensor, only the main part is converted, by the sensor itself, in a dc voltage proportional to the injected power while a residual part is lost in the sensor mount and in the feeding line, Fig.1. If the fraction of power lost in the mount is measured, the effective efficiency $h_e$ of the sensor can be obtained. Until now there are no alternatives to the calorimeter method for measuring the mount losses and a fundamental step of the process is the ability to distinguish between the power lost in the sensor mount and the power lost in the feeding line. The feeding line is the most critical component of the microcalorimeter because it must exhibit two opposing properties, that is, to be electrically lossless and thermally insulating.

*Figure 1 should be placed here.*

A thermoelectric power sensor can be supplied with a HF power $P_{HF}$ or a dc power $P_{dc}$ at which the parasitic losses, both in the feeding line and in the same sensor, may be negligible. When $P_{HF}$ and $P_{dc}$ cause the same sensor output $U$, their ratio is assumed as the effective efficiency of the sensor mount [9], that is:

$$h_e = \left. \frac{P_{dc}}{P_{HF}} \right|_{U=const.} \tag{1}$$

In Section 4 we will show that this definition, given for technical opportunity, is equivalent to a more intuitive and general form. The measurement of the effective efficiency requires, however, an appropriate correction for the losses into the feeding line. Furthermore, in order to eliminate undesired thermo-voltages that generate at each circuit interface, a low frequency (LF) power $P_{LF}$, at a frequency of 1 kHz, is used instead of the dc power $P_{dc}$, being the associated losses still negligible.

**3 – INRiM Measurement System**

At INRiM, the microcalorimeter is based on a dry thermostat [9,10] whose triple wall measurement chamber is thermally stabilized via Peltier cells and a PID controller acting on the intermediate wall. This configuration allows a thermal stability of the order of ±0.02°C measured outside the inner wall at the temperature of 25 °C while the thermostat operates in a shielded room with a temperature of $(23.0 \pm 0.3)$°C and $(50 \pm 5)$% of relative humidity. The external passive wall of the thermostat pre-filters room temperature variations while the inner one minimizes the residual temperature variations still present on the middle active wall, for PID performance limitations.
After an experience of many years on single-line inset, the twin-line configuration shown in Fig. 2 has been chosen because it behaves as full differential configuration, more effective in filtering the external thermal disturbances that still bypass the thermal shields.

*Figure 2 should be placed here.*

The main microcalorimeter detector consists in two annular arrays of Cu-Constantan thermo-junctions, designed at INRiM [10], Fig.3. These arrays measure the temperature gradient of the twin power sensors (in the following loads, also) toward the measurement environment. Their optimum position is just behind the input connectors of the same. One load (*hot-load*) is supplied alternatively with the HF/LF power, while the other (*cold-load* or *dummy-load*) is always used as thermal reference only. The two thermo-junction arrays are combined in a thermopile measuring the temperature difference between the hot and cold load.

*Figure 3 should be placed here.*

The thermopile output $e$ has an increasing exponential trend when HF power is supplied and a decreasing exponential trend when an equivalent LF power is substituted into the system. The asymptotes of these curves, which correspond to heating and cooling steps respectively, relate to the power injected in the microcalorimeter through a fundamental electro-thermal equation. This one is nothing else than the superimposition principle of the linear effects applied to our system, that is [7, 9]:

$$e = \mathbf{a}R(K_1 P_S + K_2 P_L), \qquad (2)$$

where $\mathbf{a}$ is the Seebek coefficient, $R$ is a conversion constant, $P_S$ is the total power dissipated in the sensor, $P_L$ is the total power dissipated in the feeding line, while $K_1$ and $K_2$ are separating constants. The system has two equilibrium states: one corresponds to the thermopile asymptotic value $e_1$ reached when the incoming HF power has terminated the heating step and the other, with asymptotic value $e_2$, when the system is completely cooled after the HF power is substituted with the LF power. For these two states, the thermopile outputs are:

$$\begin{cases} e_1 = \mathbf{a}R(K_1 P_S + K_2 P_L)_{HF} \\ e_2 = \mathbf{a}R(K_1 P_S + K_2 P_L)_{LF} \end{cases} \bigg|_{U=cost} \qquad (3)$$

Normally HF/LF power substitution is made maintaining the hot sensor response $U$ constant. This avoids too strong alterations of the thermodynamic equilibriums of the system, thing that can only complicate the calibration process. Combining the thermopile responses in the following ratio:

$$e_R = \frac{e_2}{e_1} = \left(\frac{P_S|_{LF}}{P_S|_{HF}}\right) \frac{1 + \frac{K_2}{K_1}\left(\frac{P_L}{P_S}\right)_{LF}}{1 + \frac{K_2}{K_1}\left(\frac{P_L}{P_S}\right)_{HF}} \bigg|_{U=const.}, \qquad (4)$$

we obtain a relation between measured quantities and effective efficiency of the hot load, that is the sensor mount under calibration for becoming the transfer standard. Assuming the definition (1) for $\mathbf{h}_e$, we obtain:

$$\mathbf{h}_e = g e_R, \qquad (5)$$

where $g$ is the microcalorimeter calibration constant, dependent on the characteristics of the feeding system. Indeed, the separation constant ratio $K_2/K_1$ is related to the thermal impedance of the insulating sections of the feeding line while the power ratios $(P_L/P_S)|_{LF,HF}$ are related to the transmission coefficient of the line in the following manner:

$$\left.\frac{P_L}{P_S}\right|_{HF,LF} = \left.\frac{1-|S_{21}|^2}{|S_{21}|^2}\right|_{HF,LF}. \quad (6)$$

Thermal parameter $K_R = K_2/K_1$, which is frequency independent, may be calculated while the power ratios may be obtained by measurements with Network Analyzers, in line of principle at least. There are problems for both, anyway. If it is very difficult to write a realistic thermodynamic model that describes the thermostat accurately, it is also difficult to decide what feeding line section really contributes to the process with its losses. A one-dimensional model was considered, based on the heat diffusion equation, for $K_R$ determination while only the feeding line section internal to the thermostat has been considered and measured for evaluating the power ratios (6). The calibration constant $g$, obtained in this way, turned out to be of poor accuracy during international comparisons among power standards. The $g$ factor should be measured directly reversing equation (5) but this requires having a hot-load of known effective efficiency, however. For all these reasons the attractive Eq. (5) is not used.

Authors proved that formula (5) reduces to a voltage ratio if the losses into the feeding line are negligible at the LF power, typically chosen at 1 kHz [10]. With this assumption (5) becomes indeed:

$$\mathbf{h}_e = \frac{e_2}{e_1 - e_{1SC}}, \quad (7)$$

Equation (7) is another experimental expression of (1), function of the microcalorimeter thermopile output $e$ only. Voltages $e_1$ and $e_2$ are the same previously specified, while $e_{1SC}$ is the response when half of the HF power generating $e_1$ is supplied to the system with the feeding line short circuited [10]. Even for this model the necessity exists of a reference thermal load of known $\mathbf{h}_e$ and how to implement such a reference load it will be the argument of the next section.

Finally, if the previous condition $U = constant$ is relaxed (this can happen when the power substitution is not perfect), by virtue of an intrinsic linearity of the system, formula (5) corrects for the voltage ratio $U_1/U_2$, being $U_1$ the sensor response to the HF power and $U_2$ that to LF power.

**4 – Implementing a reference hot-load**

The axiomatic definition of the *effective efficiency* given in Eq. (1) may be substituted by the ratio between the measured power $P_M$, i.e. the power converted in the dc output $U$ of the power sensor, and the total power absorbed by the same sensor $P_S = (P_M + P_X)$, that is:

$$\mathbf{h}_e = \frac{P_M}{P_M + P_X}, \quad (8)$$

where $P_X$ is the power loss in the sensor mount. Formula (8) can be assumed as intuitive definition of effective efficiency, but it can also be found by setting $P_S = (P_M + P_X)$ straightforwardly in (4) and noting that the HF measured power $P_{1M}$ must equal the LF measured power $P_{2M}$, if a power substitution is invoked at the condition of constant sensor output ($U = $ const.).

Now, we suppose to deal with a power sensor that from an ideal condition of perfect impedance matching (reflection coefficient $\Gamma_S = 0$) evolves with continuity into a full unmatched load ($\Gamma_S = 1$), i.e. a lossless short-circuit. According to this hypothetical transformation, measured powers ($P_M$), mount losses ($P_X$) and sensor responses ($U$) reduce to zero, but not the ratio (8) as well as the ratio $U_1/U_2$, if considered. Indeed:

$$\lim_{\Gamma_S \to 1} P_M = 0; \lim_{\Gamma_S \to 1} P_X = 0 \Rightarrow \lim_{\Gamma_S \to 1} h_e = 1$$
$$\lim_{\Gamma_S \to 1} U = 0 \Rightarrow \lim_{\Gamma_S \to 1} \frac{U_1}{U_2} = 1 \qquad (9)$$

So far, we come quickly to conclude that a perfect short circuit can be considered a load of calculable effective efficiency and therefore eligible as a reference standard for microcalorimeter calibration.

We can go further, observing that a lossless short is difficult to realize in practice, even terminating the microcalorimeter test port with a flat highly conductive surface. Typically, above 1 GHz, residual ohmic resistance worsens the reflectivity, due to the finite conductivity of the material and imperfect mechanical connections. Fortunately, we can independently determine this contribution through reflection coefficient measurements with Network Analyzers. If we apply definition (8) to a short, identifying $P_M$ with the reflected power, then the denominator of (8) coincides with the incident power and their ratio with the modulus square of the reflection coefficient of the short, i.e. $|G_{SC}|^2$.

All these important conclusions are not the results of mathematical speculations only but they are supported by an intrinsic property of the Microcalorimeter.

Indeed, definition (8) is consistent with matched power sensors as well as with unmatched loads, because the calibration process we are considering is sensitive to power losses only, independently of the reflection coefficient of its thermal load.

**5 – Power standard realization**

Effective efficiency measurement of a sensor mount is done with the following experimental sequence. The microcalorimeter feeding lines having terminated by a couple of twin mounts, we supply the hot load alternately with HF and LF power. After a significant number of repeated power substitutions, we get a record of the microcalorimeter thermometer output, as Fig. 4 shows. Measurements are taken every minute with a switching time of 90 minutes between HF and LF, about three time constants. The temperature difference between the two equilibrium states of the system is in the mK range. By means of an interpolation process, the asymptotic values $e_1$ and $e_2$ are determined, corresponding at the microcalorimeter two equilibrium states [9, 10]. The system noise may be expressed in terms of their typical statistical uncertainty that is of the order of some nV. This allows calculating the raw effective efficiency value.

*Figure 4 should be placed here.*

Then, the microcalorimeter calibration process begins by opening the thermostat and substituting the twin sensor mounts with highly reflecting twin loads of known reflection coefficient. Again, a significant number is performed of HF/LF power substitutions, obtaining another data record similar to that of Fig. 4. From this we determine two asymptotic voltages $e_{1SC}$ and $e_{2SC}$ by means of the aforementioned statistic analysis.

Finally, a corrected value of effective efficiency $h_e$ can be evaluated through equation (7) but, if the imperfection of the reflective load is considered for better accuracy, (7) changes as it follows:

$$h_e = \frac{e_2}{e_1 - e_{1SC} + \frac{e_{2SC}}{|G_{SC}|^2}}, \qquad (10)$$

Anyway the calibration step requires precautions to avoid inconsistent results. Firstly, we have to maintain the ratios between the LF and HF feeding line losses as in the case of the sensor mounts. No problem exists with the LF generator because it can maintain the same current on line independently of the load reflection coefficient. Conversely HF power must be halved by means of a self-levelling loop or by a 3 dB calibrated attenuator. In this manner the first order effects are compensated due to strong reflected waves we have with short circuit condition [9, 10].

Secondly, the reflecting loads must have the same thermal capacity, so to reproduce the initial thermal symmetry inside the thermostat. Furthermore, if the assumption is made of negligible LF losses on the feeling line, then the thermal capacity of the reflecting loads must also equal that of the sensor mount under calibration. This requirement is not simple to obtain because a power sensor is a complex inhomogeneous body, therefore its thermal parameters are difficult to calculate and even to measure. We found a solution to this problem, by inserting a shorted line section of minimum length between the microcalorimeter test ports and the power sensors. The component transforms the original absorbing load in reflective load with negligible thermal effects, realizing the electrical and thermal conditions that allow the system calibration, [10].

In summary, the *effective efficiency* of a thermoelectric power sensor is determined with two series of measurements. Firstly we supply alternatively the HF and the LF power to the sensor, to evaluate $e_1$ and $e_2$ respectively. Afterward, we feed a couple of reflective loads thermally equivalent to the original ones, alternately with appropriate HF/LF power levels in order to determine $e_{1SC}$ and $e_{2SC}$. The four asymptotic values are evaluated with a fitting procedure and an example of fitted curve can be found in [9] while the reflection coefficient of the short circuit $G_{SC}$ is measured with a VNA.

## 6 – Experimental results

To show well the importance of the microcalorimeter calibration, i.e. the error correction for the feeding line losses, we present the results concerning primary transfer standard calibration based on indirect heating thermoelectric sensor. The standard has been obtained by using three mathematical models and Fig. 5 shows how the *effective efficiency* values are related to these models. Thin line represents the raw $h_e$ calculated with the relation:

$$h_e = \frac{e_2}{e_1} \qquad (11)$$

which could be considered relevant to a lossless system. Solid line is relevant to the model (7) which accounts for the losses on the feeding lines but still does not include the losses in the reference load. Finally, dashed line includes also the correction related to the real short circuit used to calibrate the microcalorimeter (Eq. (10)).

As expected, only the correction related to the feeding line losses is important, while the term relative to the reflection coefficient of the short circuit becomes important only beyond 20 GHz, where the short circuit performances degrade significantly, as Fig.6 shows. A correction for the power substitution error, that is when $U_1/U_2 \neq 1$, has also been considered but, in our case, it resulted negligible because we were able to obtain very good power control. Indeed the curve (dash-dot line) overlaps very well the curve relative to the short circuit correction (dashed line).

*Figure 5 should be placed here.*

*Figure 6 should be placed here.*

The uncertainty on the presented results has been evaluated, after a measurement period lasted about four months, through a standard Gaussian propagation, following the recommendation of the GUM [11]. The uncertainty components are the ones related to $e_1$, $e_2$, $e_{1SC}$, $e_{2SC}$, coming from fitting procedures, to $G_{SC}$ coming from VNA measurements and also to $U_1$ and $U_2$ coming from repeated measurements of the sensor output voltage.

Figure 7 shows the total relative uncertainty on $h_e$ obtained on the frequency range from 10 MHz to 26.5 GHz. The values are reported with a coverage factor $k = 2$ and are well under 0.45 % in all the measurement range. The dashed line is a guide for the eye and shows that the majority of the points are below 0.25% in all the considered frequency range.

*Figure 7 should be placed here.*

These uncertainty figures are significant if compared to the ones generally reported in literature. Indeed, even in international key comparisons, values below 1% are not usual, at least for the highest frequencies of the considered frequency band [12].

**7 – Conclusions**

This work is aimed to describe the details of the microcalorimetric technique. Despite the technique dates back to the first 50s, it is still confined in the laboratories that provide the national standards and no commercial realization exists, indeed, with detriment to the same technique. Therefore the highlights are on the physical principles that are at the base of the technique and on the derivation of the mathematical modes that describe the system behaviour at the best. We gives a model that can account for the main error sources, supposing negligible the LF losses and the non-linearity contribution both of the thermopile and power sensor. As this model is not easy to use in practice, despite the appearance, we mention also simplified models that we were able to propose in previous referenced works. Experimental data are reported also, relevant to a comparison between different microcalorimeter models and their analysis supports the everywhere consolidated assumption that only the feeding line losses are a huge error source.

Most important hint concerns the reference load used to calibrate the system. We explain why a short circuit may be assumed as calculable standard of effective efficiency, despite its huge electrical difference from a power sensor.

The exposed theory applies to thermoelectric detectors because, for opportunity reasons, we prefer avoiding bolometric sensors. Anyway it may be adjusted to these ones, with little efforts.

**References**

[1]   J. D. Kraus, *Electromagnetics,* McGraw-Hill, Inc., International Edition 1991.
[2]   A. C. MacPherson, D. M. Kerns, *Rev. Sci. Instrum.*, **26**, pp. 27-33, 1955.
[3]   G. F. Engen, *J. Res. Nat. Bur. Stand.*, **63C**, pp. 79-82, 1959.
[4]   R. F. Clark, *Proc. IEEE*, **74**, 1, pp. 102-104, Jan. 1986.
[5]   N. S. Chung, J. Shin, H. Bayer, R. Honigbaum, IEEE Trans. Instrum. Meas., **38**, 2, pp. 460-464, 1989.
[6]   T. W. Kang, N. S. Chung, R. Honigbaum, J. Rühaak, U. Stumper, IEEE Trans. Instrum. Meas., **46**, 6, pp. 1247-1250, 1997.
[7]   A. Fantom, *Radiofrequency & microwave power measurement*, Peter Peregrinus Ltd., England, 1990.

[8]  E. Völlomer, et al., in *Proc. CPEM'94*, pp.147-148, 1994.
[9]  L. Brunetti, E. Vremera, IEEE Trans. Instrum. Meas., **52**, 2, pp. 320-323, April 2003.
[10] L. Brunetti, L. Oberto, E. Vremera, IEEE Trans. Instrum. Meas., **56**, 6, pp. 2220-2224, December 2007.
[11] BIPM, IEC, IFCC, ISO, IUPAC, IUPAP and OIML, *Guide to the Expression of the Uncertainty in Measurenent*, 2$^{nd}$ edn., 1995, ISBN 92-67-10188-9.
[12] D. Janik, et al., Metrologia, **43**, Technical Supplement, 01009, 2006.

FIGURE CAPTIONS:

Fig. 1: Microcalorimeter scheme in terms of main components. $P_S$ represents the total absorbed power by the sensor that includes the power really measured and the parasitic losses in the sensor mount.

Fig. 2: Pictorial view of the twin-line coaxial Microcalorimeter whose inset is composed by a thermopile assembly, insulating line sections and two identical sensor mounts.

Fig. 3: Thermopile assemblies for twin coaxial microcalorimeter.

Fig. 4: Example of raw data recorded at the frequency of 26 GHz.

Fig. 5: Effective efficiency trends. Thin line – raw values; solid line – correction for line losses only; dashed line – correction for line losses and for non ideal reflecting load; dash-dot line – all corrections.

Fig. 6: Reflection coefficient trend of the reflecting load used to calibrate the microcalorimeter.

Fig. 7: Total extended relative uncertainty on the effective efficiency of an INRiM transfer standard based on thermoelectric sensor.

FIGURES:

Figure 1

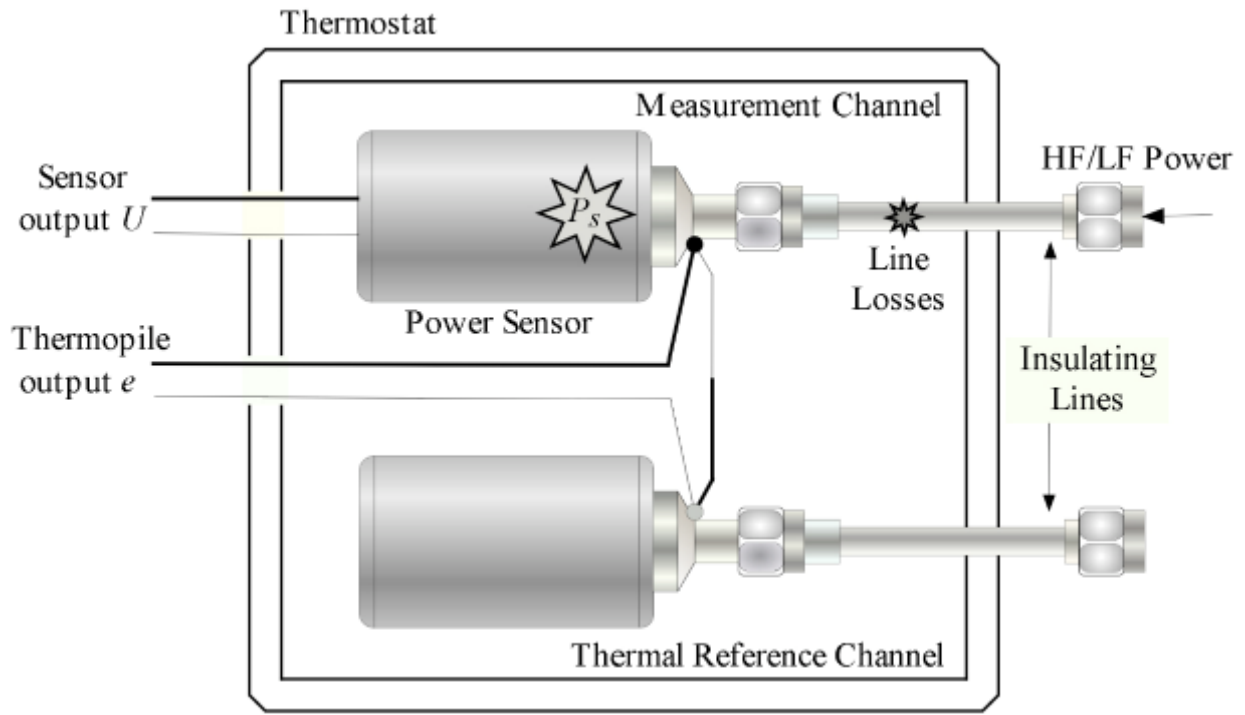

Figure 2

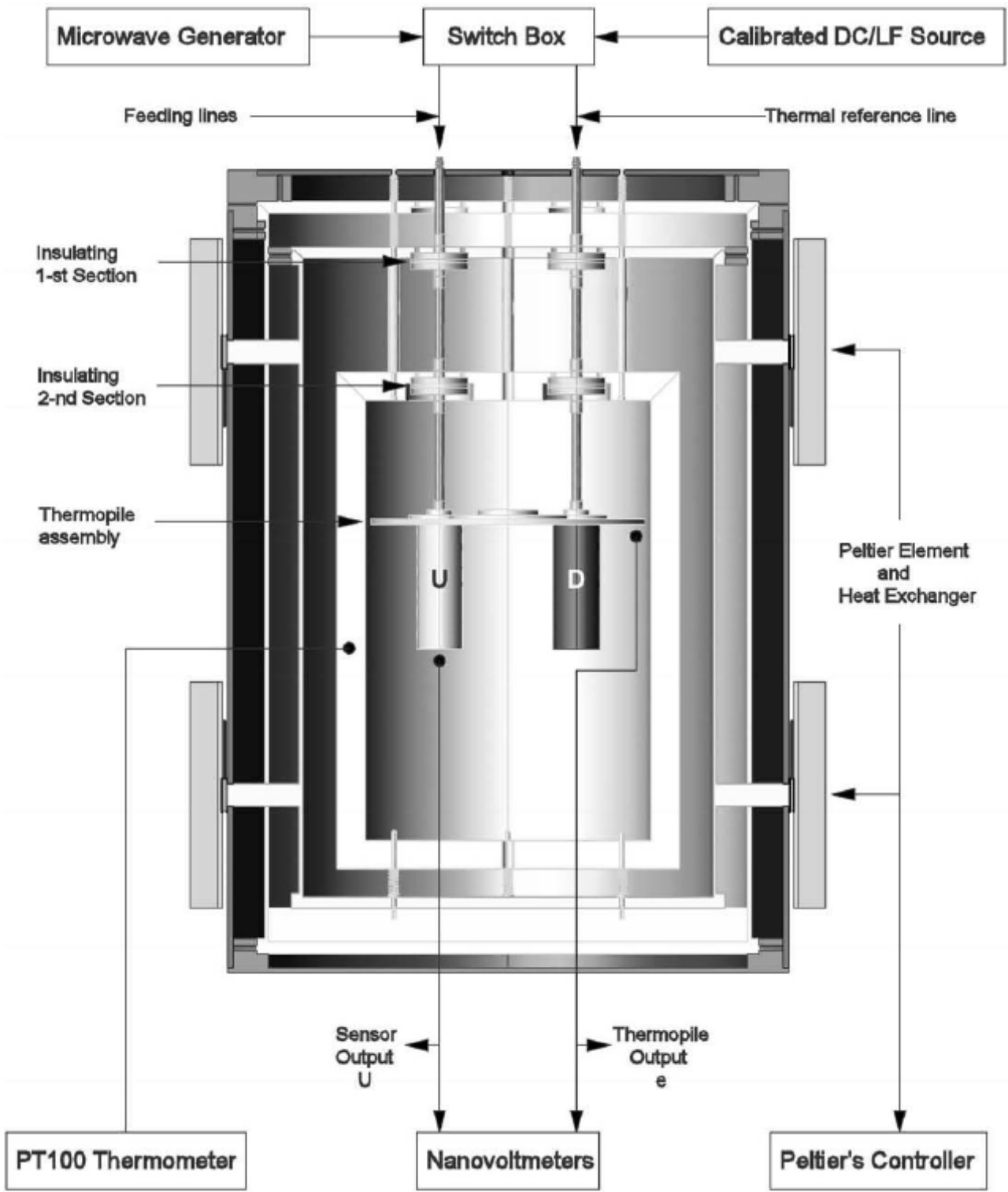

Figure 3

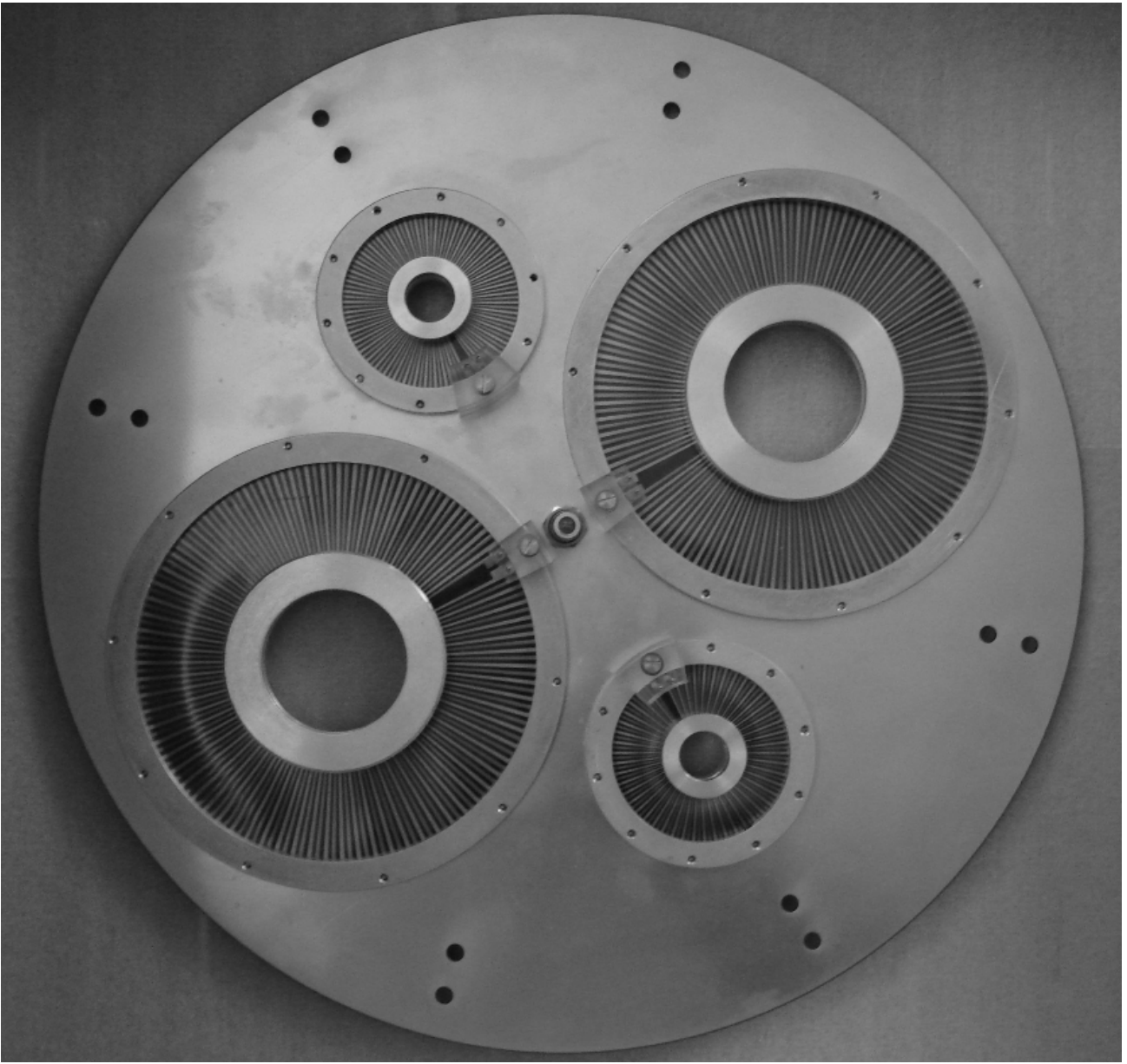

Figure 4

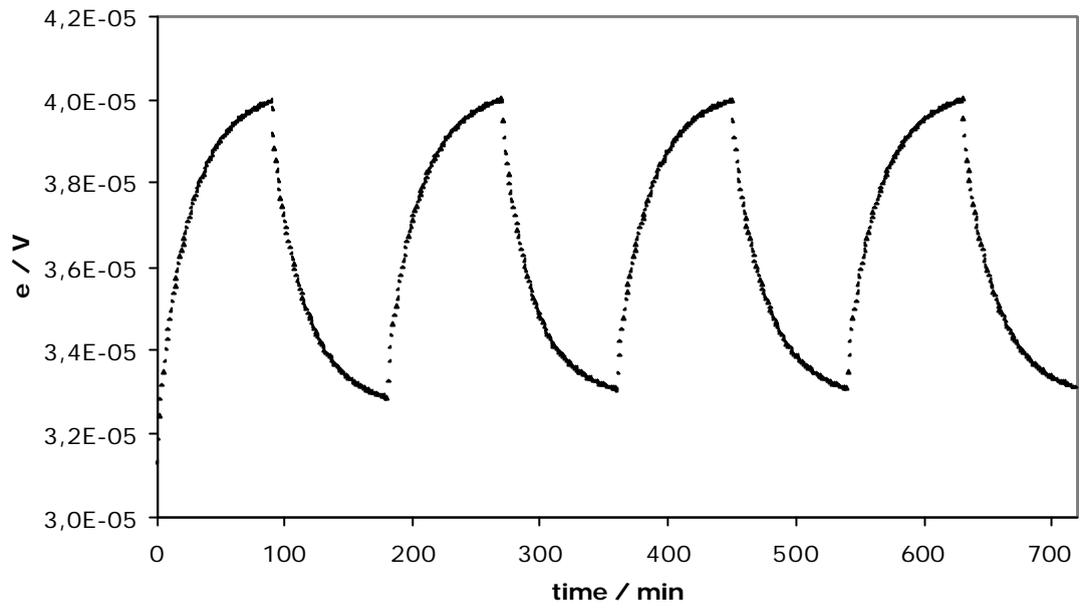

Figure 5

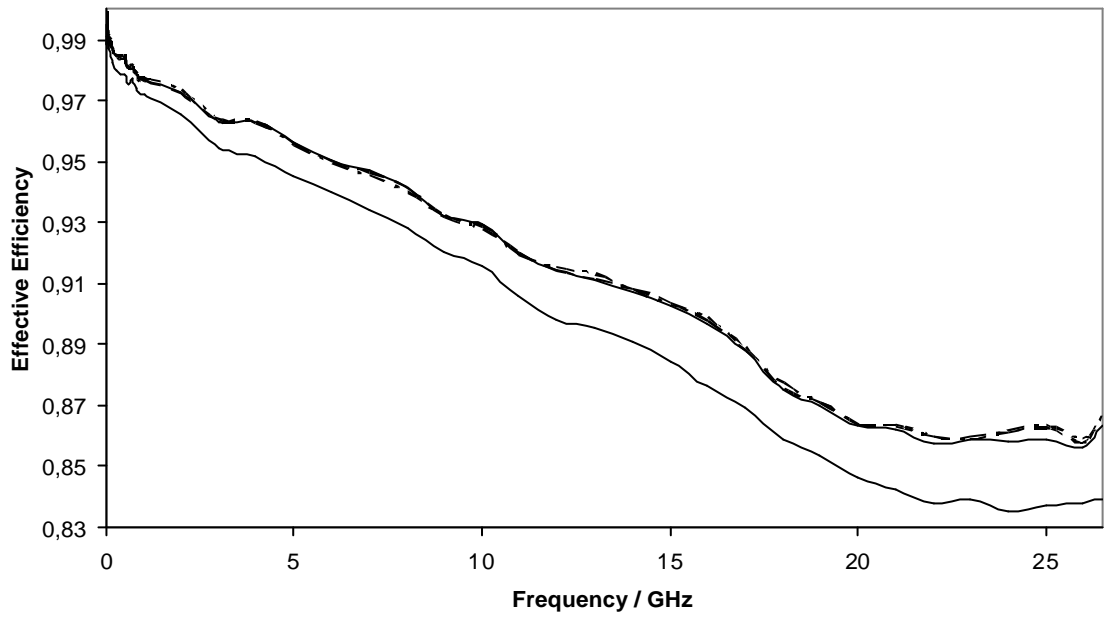

Figure 6

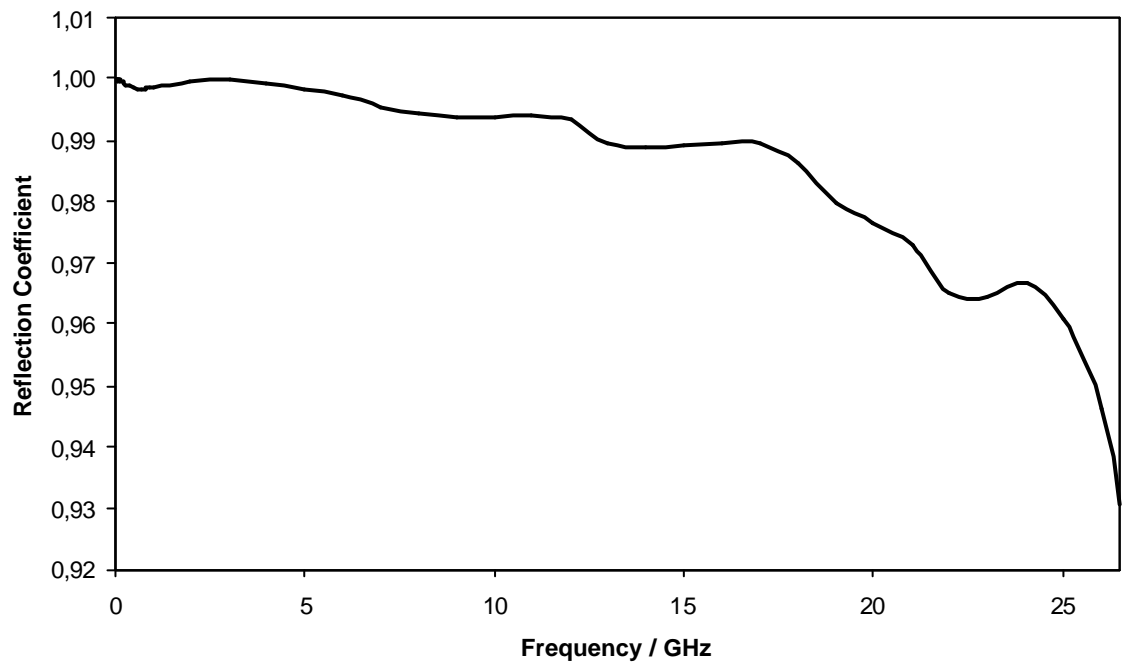

Figure 7

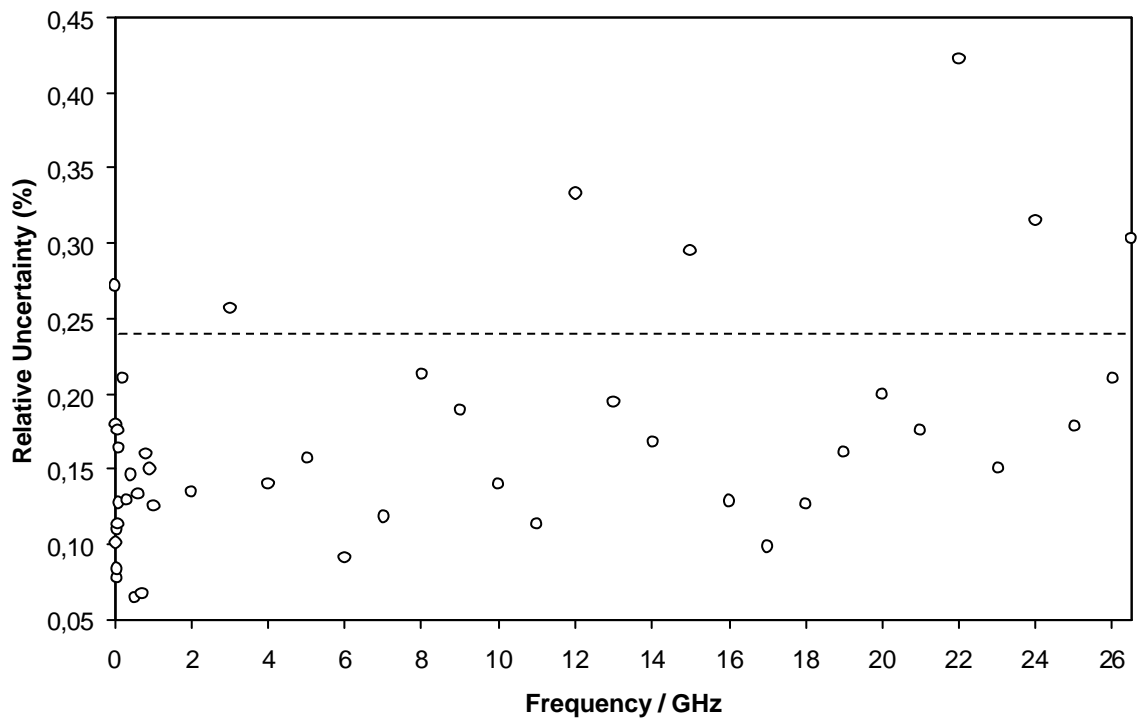